\documentclass[letter]{IEEEtran}
\usepackage{cite}
\usepackage[cmex10]{amsmath}
\usepackage{graphics}
\usepackage{graphicx}
\usepackage{mathtools,amsbsy,bbm,color}
\usepackage{amssymb}
\usepackage{float}
\usepackage[font=small, singlelinecheck=false]{caption}
\usepackage{caption}
\usepackage[caption = false]{subfig}
\usepackage{fancyhdr}

\usepackage{url}
\usepackage{xr}
\allowdisplaybreaks

\newtheorem{lemma}{Lemma}

\newenvironment{proof}[1][Proof:]{\begin{trivlist}
\item[\hskip \labelsep {\bfseries #1}]}{\end{trivlist}}

\AtBeginDocument{%
 \abovedisplayskip = 0.04in
 \abovedisplayshortskip = 0.04in
 \belowdisplayskip =  0.04in
 \belowdisplayshortskip = 0.04in
 \topsep = 0.04in
 \textfloatsep = 0.05in
 \textheight = 9.35in
}
\begin{document}

\title{\hspace{-0.0in}Optimizing Mission Critical Data Dissemination in Massive IoT Networks
%Improving Spatial Reuse and Packet Dissemination in Multi-hop CSMA Networks
}

\author{ \IEEEauthorblockN{\large Muhammad Junaid Farooq$^*$, Hesham ElSawy$^\dag$, Quanyan Zhu$^*$, and Mohamed-Slim Alouini$^\dag$,}\\
\IEEEauthorblockA{$^*$Department of Electrical \& Computer Engineering, Tandon School of Engineering, \\ New York University, Brooklyn, NY, USA,  \\$^\dag$King Abdullah University of Science and Technology (KAUST), Thuwal, Saudi Arabia,\\
Emails: \{mjf514,qz494\}@nyu.edu, \{hesham.elsawy, slim.alouini\}@kaust.edu.sa.}\vspace{-0.2in}}

%\\ Email: \{muhammadjunaid.farooq, hesham.elsawy, slim.alouini\}@kaust.edu.sa.\vspace{-0.4in}} }
%\author{ \IEEEauthorblockN{\large Author 1, Author 2, and Author 3}\vspace{-0.3in} }
%\thanks{Muhammad~Junaid~Farooq, Hesham~ElSawy and Mohamed-Slim~Alouini are with the Computer, Electrical, and Mathematical Sciences and Engineering (CEMSE) Division at King Abdullah University of Science and Technology (KAUST), Thuwal, Makkah Province, Kingdom of Saudi Arabia (e-mail:~\{muhammadjunaid.farooq, hesham.elsawy, slim.alouini\}@kaust.edu.sa).}
\maketitle
\IEEEpeerreviewmaketitle

\begin{abstract}
Mission critical data dissemination in massive Internet of things (IoT) networks imposes constraints on the message transfer delay between devices. Due to low power and communication range of IoT devices, data is foreseen to be relayed over multiple device-to-device (D2D) links before reaching the destination. The coexistence of a massive number of IoT devices poses a challenge in maximizing the successful transmission capacity of the overall network alongside reducing the multi-hop transmission delay in order to support mission critical applications. There is a delicate interplay between the carrier sensing threshold of the contention based medium access protocol and the choice of packet forwarding strategy selected at each hop by the devices. The fundamental problem in optimizing the performance of such networks is to balance the tradeoff between conflicting performance objectives such as the spatial frequency reuse, transmission quality, and packet progress towards the destination. In this paper, we use a stochastic geometry approach to quantify the performance of multi-hop massive IoT networks in terms of the spatial frequency reuse and the transmission quality under different packet forwarding schemes. We also develop a comprehensive performance metric that can be used to optimize the system to achieve the best performance. The results can be used to select the best forwarding scheme and tune the carrier sensing threshold to optimize the performance of the network according to the delay constraints and transmission quality requirements.
\vspace{-0.1in}
\end{abstract}
%\begin{keywords}
%stochastic geometry, spatial reuse, CSMA, PPP, MHCPP.
%\end{keywords}

\section{Introduction}
The Internet of things (IoT) is enabling an ever increasing number of services and applications that is revolutionizing the way we interact with our surroundings~\cite{iot_survey}. The ability to remotely monitor and manage objects in the physical world is leading to a paradigm shift in sectors such as transportation, healthcare, public safety, energy management, home and industrial automation, etc.~\cite{iot_ref1}. The future of IoT lies in ubiquitous sensing and connected systems that will provide enhanced situational awareness, data driven decision analytics, and automated response without human intervention~\cite{iot_ref2}. The number of IoT devices\footnote{We refer to both sensors and actuators as IoT devices.} is growing at an increasing pace leading towards what is being referred to as the massive IoT~\cite{massive_iot}. It is estimated that the number of IoT devices will reach $1.5$ billion in $2022$~\cite{number_iot}. There is a wide variety of data carried over IoT networks, some of which might be mission critical such as in applications that require real-time monitoring and control. Hence, the successful transport of data with minimum delay from the source to destination is highly essential for effective operation of massive IoT systems.

Despite the fact that most IoT networks are infrastructure based, i.e., the sensors and actuators communicate with and are controlled by access points (APs) or base stations, the connectivity in these networks relies heavily on the use of device-to-device (D2D)~\cite{d2d_iot} or machine-to-machine (M2M)~\cite{M2M} communications. This is mainly due to the following two reasons: (i) The low power remotely deployed IoT devices have limited communication range; therefore, the sensor data and control commands may have to traverse several D2D links to reach one of the APs. (ii) In certain application scenarios such as in geographically remote areas or battlefields~\cite{junaid_wiopt}, the traditional communication infrastructure such as cellular networks might be unavailable, which necessitates the use of D2D communication. Furthermore, simultaneous transmission by the massive number of devices requires medium access control (MAC) to regulate limited spectrum access. Contention based medium access protocols such as the carrier sense multiple access (CSMA) used for WiFi transmissions is well known to improve performance alongside providing equal opportunity of transmission to each device in the long run. Although efficient MAC design for IoT is still an active area of research, the contention based schemes are the most practical and viable ones~\cite{csma_iot}. In fact, many existing IoT devices either use traditional WiFi or are designed based on the IEEE 802.11 MAC protocol.

Contention based MAC protocols coordinate spectrum access among neighbouring devices to allow only non-overlapping spatial transmissions to alleviate dominant interferers. The number of simultaneous transmitters in the network is controlled by a critical design parameter referred to as the carrier sensing threshold (CST) or detection threshold~\cite{CSMA_threshold}. Increasing the CST enables more aggressive reuse of the spectrum over the spatial domain as more devices can transmit simultaneously. However, it also increases the interference which degrades the chances of transmission success. Hence, the CST imposes a delicate tradeoff between the \emph{transmission quality}, measured by the probability of transmission success, and  the \emph{spatial frequency reuse} (SFR), quantified by the spatial density of concurrent transmitters \cite{spatial_reuse}. This performance tradeoff has been well recognized in literature and attempts have been made to maximize the spatial density of successful transmissions by tuning the CST~\cite{Baccelli_vol_2}.

The efficient data dissemination in the case of mission critical massive IoT networks imposes additional performance objectives beyond maximizing the density of successful transmissions. In mission critical applications, the message transfer delay is crucial which requires the introduction of packet progress into the performance metric~\cite{mission_critical_mac}. Although a metric known as the \emph{density of progress}, which measures the mean number of meters progressed by all transmissions per unit area, has been introduced for the case of ALOHA medium access~\cite{Baccelli_vol_2}, it does not guarantee multi-hop progress from the source towards a particular destination. Moreover, the inherent intractability in analyzing the performance of CSMA inhibits further intricate analysis and hence, the notion of packet progress has not been investigated in this case. For CSMA networks,  most analytical results in literature are presented for fixed transmitter-receiver distance in a single hop setting~\cite{alfano_dense_csma,hesham_modified_hcpp,Baccelli_csma}. The model in~\cite{junaid_vehicular,renzo} studies multi-hop CSMA for vehicular networks; however, the developed models are only suited for one dimensional scenarios.

Transmitting data over multiple hops from source to destination requires selection of relay devices at each hop. In general, there can be numerous strategic ways to select potential forwarding devices among the neighbours; however, we investigate the extreme cases as used in~\cite{nardelli}, whereby the devices select the relay which results in the most progress towards the destination or the device which is the nearest to the transmitter. It helps in establishing meaningful performance bounds for the system. These forwarding strategies are referred to as the \emph{Most Forward with Fixed Radius (MFR)} and \emph{Nearest with Forward Progress (NFP)} respectively. In addition, we also study the intermediate benchmark case where a relay is selected at random from one of the available forward neighbours. This strategy is referred to as \emph{Random with Forward Progress (RFP)}. The \emph{MFR} strategy results in minimum number of hops to reach the destination, however, it may require multiple retransmissions due to low success probability that emanates from the distance dependent power decay. On the contrary, \emph{NFP} strategy may require more number of hops to reach the destination but less retransmissions due to the high success probability resulting from shorter hop distance. The \emph{RFP} may have a mix of long and short hops resulting in a mid point performance between the \emph{MFR} and \emph{NFP}. It is important to note that there is an interplay between the packet forwarding scheme and the CST. For instance, to maintain the same success probability, a forwarding scheme that relies on longer hops requires more conservative CST than the scheme that relies on shorter hops. Therefore, it is imperative to develop a comprehensive performance metric that can accurately capture these tradeoffs to optimize mission critical data dissemination in massive IoT networks.

In this paper, we exploit tools from stochastic geometry to model the system performance under the aforementioned packet forwarding schemes, namely the \emph{MFR}, \emph{NFP}, and \emph{RFP}. The main performance parameters are the transmission success probability, the density of concurrent  transmitters, and the forward packet progress. To simultaneously capture the effect of these three inter-related parameters, we use the average packet progress (APP) metric defined as the average successful progress made by packets towards the destination per unit area. The results show that the choice of forwarding scheme and the carrier sensing threshold that maximizes the average packet progress depends on the delay tolerance and the transmission quality requirements.

%Transmitting to the nearest neighbour can lead to a higher quality of transmission due to a stronger channel but at the same time will lead to a smaller geographical progress towards the destination. On the contrary, transmitting to the farthest available relay results in lower quality of transmission but with higher progress towards the destination. Transmitting to a random neighbour will result in a midpoint performance between the two extremes. The progress determines the number of hops required to reach the destination and hence affects the transmission time from the source to the destination.

%The spatial reuse is controlled by the carrier sensing threshold only while the forwarding scheme affects both the spatial reuse and the packet dissemination performance in terms of the transmission delay.

%In this letter, we consider a random ad-hoc network using a single carrier frequency. We use a stochastic geometry approach to quantify the performance of the multihop ad-hoc network under the three packet forwarding schemes. A novel performance metric has been developed, called the aggregate packet progress, that captures the spatial reuse and the efficiency of packet dissemination in the network.

%The density of progress as mentioned in Section 16.3 is used to measure the mean number of meters progressed by all transmissions. However, it is measures generalized progress for a single hop does not give meaningful insight on the actual directional progress or the delay to reach the destination and cannot be generalized for multi-hop communication.

\vspace{-0.1in}
\section{System Model}
\begin{figure}[t]
  \centering
  \includegraphics[width=2.5in]{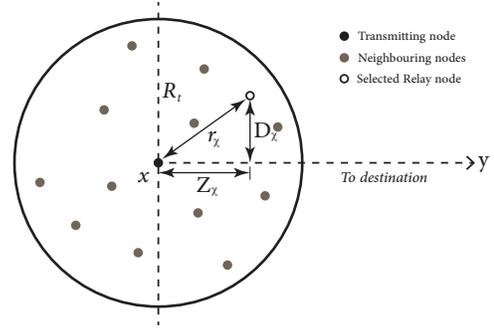}
  \caption{Packet forwarding at each hop. The transmitter node at location \emph{x} forwards the packet to one of its forward neighbours based on the transmission scheme. The forward progress is the distance covered by the packet towards the destination node at location \emph{y}.}\label{system_model}
\end{figure}

We consider a network of uniformly scattered IoT devices in $\mathbb{R}^2$ represented by a marked homogeneous Poisson Point Process (PPP) $\Phi = \{ X_{i},M_{i},T_{i} \}_{i \geq 1}$ of intensity $\lambda$, where the tuple $ \{X_{i}, M_i, T_i\}$ denotes, respectively, the location, i.e., spatial coordinates, mark, and status of the  $i^{th}$ device. The marks $M_{i} \in [0,1]$ are assumed to be independent and identically distributed (i.i.d.) uniform random variables, referring to the backoff timer value in the contention-based CSMA MAC protocol. The status $T_{i}$ is a Boolean value, which is $1$ if the device at $X_{i}$ transmits and is $0$ otherwise. According to the contention process, $T_{i}$ will be $1$ only if the $i^{th}$ node has the least mark in its average contention range denoted by $\bar{R}_s$, i.e., $ M_i = \underset{j}{\min} \{M_j:  \|X_{i} - X_{j}\| \leq \bar{R}_{s}, j \geq 1\}$, where $\|.\|$ is the Euclidean norm.

We assume that all devices transmit with a fixed transmit power $P$ and the power decays with the distance $r$ according to the \emph{power law} $r^{-\eta}$, where $\eta>2$ is the path loss exponent. The power received from a transmitting device at $X_{i}$ by the receiver at $X_{j}$ is therefore evaluated as $P h_{ij} \|X_{i}-X_{j}\|^{-\eta}, i \neq j$, where $h_{ij}$ is the random channel fading gain between locations $X_{i}$ and $X_{j}$. We assume that the channel gains $h_{ij}$ are i.i.d. exponential random variables with mean $\mu^{-1}$. According to the \emph{block fading} channel model, the channel gains are assumed fixed for the duration of each time slot and independent of the other time slots. The transmission range of the transmitter denoted by $R_{t}$ is defined as $R_{t} = \{ \|X_{i} - X_{j} \|  :Ph_{ij} \|X_{i} - X_{j} \|^{-\eta} \geq \rho_{min} , \forall i,j \geq 1\}$, where $\rho_{min}$ is the minimum signal power for detection. The average transmission range can be expressed as $\bar{R}_t = \left( \frac{P}{\mu \rho_{\min}}\right)^{\frac{1}{\eta}} \Gamma \left(1 + \frac{1}{\eta}\right)$. Similarly, the contention range $R_{s}$ is defined as $R_{s} = \{ \|X_{i} - X_{j} \|  :Ph_{ij} \|X_{i} - X_{j} \|^{-\eta} \geq \rho_{th}, \forall i,j \geq 1 \}$, where $\rho_{th}$ is the CST, which is a central design parameter in this paper. The average contention range can be expressed as $\bar{R}_{s} = \left( \frac{P}{\mu \rho_{th}}\right)^{\frac{1}{\eta}} \Gamma \left(1 + \frac{1}{\eta}\right)$.

Fig.~\ref{system_model} illustrates an example realization of the network with uniformly deployed devices in $\mathbb{R}^2$. Note that only the selected transmitting device located at \emph{x} and its neighbors inside the average transmission range are shown. The data originates from the device placed at \emph{x} to a device or access point located at \emph{y} arbitrarily far from the transmitter. The transmitter at \emph{x} selects one of its available neighbors as a relay and forwards the data. The choice of next hop relay is determined by the forwarding scheme and is restricted to the \emph{forward} direction only, i.e., towards the destination. The per-hop transmitter-receiver separation, denoted by $r_{\chi}$, is a random variable and depends on the adopted forwarding scheme $\chi$, where $\chi \in \{ M,N,R\}$ referring to the \emph{MFR}, \emph{NFP} and \emph{RFP} strategies respectively. At each hop, a similar situation is encountered where the relay becomes the transmitter and hence, it is sufficient to analyze single hop performance.

We assume saturation conditions in which all devices always have backlogged data in their buffers, and hence, all of them will contend for spectrum access in each time slot\footnote{Extension to unsaturated buffers is straightforward and is dealt in \cite{VTC}.}.
%Note that this simplifying assumption does not change the insights of the results as shown in \cite{VTC}.
At a given time slot, only the nodes that are qualified under the contention based MAC, i.e., $\{ X_{i} \in \Phi: T_{i} = 1 \}$. The value of $T_{i}$ is determined by the contention domain of $X_{i}$, i.e., $\mathcal{N}_{i} = \{ X_{j} : P h_{ij} \|X_{j} - X_{i} \|^{-\eta} \geq \rho_{th}, \forall j \geq 1, j \neq i\}$. It is clear that $\rho_{th}$ controls the cardinality of the contention domain denoted by $\mathbb{E}|\mathcal{N}_{i}|$ for all $X_{i} \in \Phi$. The set of concurrent transmitting nodes selected by the CSMA protocol is denoted by $\tilde{\Phi} = \{ X_{i} \in \Phi : T_{i} = 1 \}$, and is modeled using a Mat\'ern hard core point process (MHCPP). The MHCPP is obtained by a dependent thinning of the PPP $\Phi$ to obtain $\tilde{\Phi} \subseteq \Phi$. The resulting non-homogeneous PPP can be reasonably approximated by a homogeneous PPP with the following intensity~\cite{mean_interference_haenggi}:
\begin{equation}
\Lambda(\rho_{th}) = \lambda \frac{1 - e^{-\mathbb{E}|\mathcal{N}_{i}|}}{\mathbb{E}|\mathcal{N}_{i}|},
\end{equation}
%where $p$ is known as the \emph{retaining} probability and is calculated as follows:
%\begin{equation}
%p(\rho_{th}) = \frac{1 - e^{-\lambda \mathbb{E}|\mathcal{N}_{i}|}}{\lambda \mathbb{E}|\mathcal{N}_{i}|}
%\end{equation}
where $\mathbb{E}|\mathcal{N}_{i}|$ is the expected number of nodes in the contention domain of $X_{i}$. In our case, it is equivalent to the device density times the area of the contention domain i.e. $\mathbb{E}|\mathcal{N}_{i}| = \lambda \pi \bar{R}_{s}^{2}, \forall i\geq 1$. Note that $\mathbb{E}|\mathcal{N}_{i}|$ is a function of $\bar{R}_{s}$ which in turn is a function of the CST $\rho_{th}$.

\vspace{-0.1in}
\section{Analysis}
In this section, we first characterise the transmission quality at each hop in terms of the probability of transmission success for a fixed transmitter-receiver distance. We then provide the distribution of distances in the three packet forwarding strategies. Finally, we evaluate the distributions for the forward packet progress achieved by using each of the forwarding strategies.
\subsection{SINR Characterization}
The successful data delivery between the transmitter and receiver is dependent on the signal-to-noise-plus-interference-ratio ({\rm SINR}). We use a \emph{signal capture} model at the receiver with a detection threshold $\beta$, i.e., a transmission is successful or correctly decoded by the receiver only if the {\textrm SINR} is above the detection threshold $\beta$, which signifies the quality of service (QoS) requirement of the network. Without loss of generality, we can consider the receiving node to lie on (0,0) in $\mathbb{R}^{2}$. Since the per-hop transmitter-receiver separation distance depends on the forwarding strategy $\chi$, each one of the strategies has a different {\textrm SINR} that is described as follows:
\begin{equation}
{\rm SINR}_{\chi} = \frac{P h_{i0} r_{\chi}^{-\eta}}{\kappa + I_{\tilde{\Phi} \backslash X_{i}}},
\end{equation}
where $\kappa$ is the thermal noise power and $I_{\tilde{\Phi}  \backslash X_{i}}$ is the \emph{aggregate interference} power emanating from the set of simultaneous transmitters excluding the tagged transmitter $X_i$, defined as $I_{\tilde{\Phi} \backslash X_{i}} = \sum \limits_{j = \{k \geq 1 : X_{k} \in \tilde{\Phi} \backslash X_{i}\}} P h_{j0} \|X_{j}-0\|^{-\eta}$. The probability of transmission success for the forwarding strategy $\chi$ can be expressed as follows:\vspace{-0.2in}

\begin{align}
\small
\mathcal{P}_{\chi}&= \mathbb{P}[SINR > \beta] = \int_r \mathbb{P}\left[\frac{Ph_{i0} r^{-\eta}}{\kappa+I_{\tilde{\Phi} \backslash X_{i}}}>\beta\right] f_{r_{\chi}}(r) dr, \notag \\
%&\overset{(a)}{=}  \int_r \exp\left\{- \frac{T \mu \kappa r^{\eta}}{P}\right\}  \exp\left\{- \frac{T \mu I_{agg} r^{\eta}}{P}\right\} f_{r_o}(r) dr \notag \\
&=  \int_r \exp\left\{- \frac{\mu \beta \kappa r^{\eta}}{P}\right\} \mathcal{L}_{I_{\tilde{\Phi} \backslash X_{i}}}\left(\frac{ \mu \beta r^{\eta}}{P}\right) f_{r_{\chi}}(r) dr,
\label{P_success_LT}
\end{align}

\noindent where $\chi \in \{ M,N,R\}$. $\mathcal{L}(.)$ is the Laplace transform (LT) of the probability density function ({\em pdf}) of $I_{\tilde{\Phi} \backslash X_{i}}$. From (\ref{P_success_LT}), it is clear that we have to obtain the LT of the aggregate interference as well as the {\em pdf} of $r_{\chi}$ in order to derive the probability of successful transmission. For tractability of analysis, the interference seen from the MHCPP $\tilde{\Phi}$ can be approximated by the interference seen from an equi-dense PPP \cite{hesham_modified_hcpp}. Therefore, the LT of of aggregate interference is derived as follows:
\begin{align}
\mathcal{L}_{I_{\tilde{\Phi} \backslash X_{i}}} (s) &= \mathbb{E}\left[e^{-sI_{\tilde{\Phi} \backslash X_{i}}}\right] = \mathbb{E} \left[e^{-s \sum \limits_{j = \{k : X_{k} \in \tilde{\Phi} \backslash X_{i}\}} P h_{j0} v^{-\eta}} \right], \notag\\
&= \exp \left( - 2 \pi \Lambda \int \limits_{\phi \in \mathbb{R}^{2}} \left( 1 - \mathcal{L}_{h_{j0}} (sPv^{-\eta}) \right) v dv \right), \label{phi_eq} \\
&\geq \exp \left( - 2 \pi \Lambda \int \limits_{\bar{R}_{s} - r}^{\infty}  \frac{1}{1 + \frac{\mu}{P s v^{-\eta}}} v dv \right),
\label{LT}
\end{align}
Note that integrating over the exact interference region denoted by $\phi$ in \eqref{phi_eq} is not possible in closed form due to its geometry with respect to the origin. However, observing that the nearest interferer can be located at a distance of $\bar{R}_s - r$ from the origin, a lower bound can be obtained by using the overestimated interference region as $\phi= \mathbb{R}^{2} \setminus b\left(0,\bar{R}_{s} - r_{\chi}\right)$, where $b(x,y)$ is a ball of radius $y$ centered at $x$. The lower bound of the LT in \eqref{LT} can be expressed in closed form for $\eta = 4$. In this case, the probability of transmission success can be obtained by numerically computing the following integral:
\begin{equation}
\mathcal{P}_{\chi} \stackrel{(\eta = 4)}{=} \int \limits_{0}^{\bar{R}_{t}} e^{- \frac{\mu \beta \kappa r^{4}}{P}} e^{ - \pi \Lambda \sqrt{\beta} r^{2}  {\rm tan}^{-1} \left(  \frac{\sqrt{\beta}r^{2}}{(\bar{R}_{s}- r)^{2}} \right) } f_{r_{\chi}}(r) dr.
\end{equation}
In the following subsection, we provide the \emph{pdf}s of the transmitter-receiver distance at each hop for the three main packet forwarding strategies, which is required to compute the probability of successful transmission.\vspace{-0.1in}
\subsection{Distance Distribution}
The distance between transmitter and receiver at each hop depends on the choice of selected relaying devices at each hop. For the \emph{MFR} forwarding scheme, the transmitter-receiver distance is denoted by $r_{M}$ and its distribution can be expressed by the following lemma:
\begin{lemma} \label{mfr_dist}
The distance between the transmitter and selected relay that maximizes the forward progress within the average transmission range $\bar{R}_t$ can be expressed as follows:
\begin{align}
\small
f_{r_{M}}(r) &= \hspace{-0.1in}\int \limits_{-\frac{\pi}{2}}^{\frac{\pi}{2}} \frac{\lambda r e^{  -\lambda \bar{R}_t^{2} \left(cos^{-1}[ \frac{r}{\bar{R}_t}cos \theta] - \frac{r}{\bar{R}_t}cos \theta \sqrt{1-(\frac{r}{\bar{R}_t}cos \theta)^{2}}  \right) }   }{1 - e^{-\lambda \pi \bar{R}_{t}^{2}/2}} d \theta, \notag \\
&\hspace{2.in}0 \leq r \leq \bar{R}_t.
\end{align}
\normalsize
\begin{proof}
See \textbf{Appendix~\ref{proof_mfr_dist}}.
\end{proof}
\end{lemma}
For the {\em NFP} forwarding strategy, the transmitter-receiver distance is denoted by $r_{N}$ and its distribution can be expressed by the following lemma:
\begin{lemma}
The distance between the transmitter and the nearest available relay device in the average transmission range $\bar{R}_t$ can be expressed as follows:
\begin{equation}
f_{r_{N}} (r) = \frac{\lambda \pi r e^{-\lambda \pi r^{2}}}{1 - e^{-\lambda \pi \bar{R}_{t}^{2}/2}}, \qquad \hfill{0 \leq r \leq \bar{R}_{t}.}
\label{f_r_N}
\end{equation}
\begin{proof}
The distribution of distance to the nearest neighbour in Poisson networks is provided in~\cite{distances}. However, note that the \emph{pdf} of the nearest neighbour distance has been conditioned on the probability that there is at least one available node in the forward neighbourhood of the transmitter.
%For the \emph{pdf} of the distance between the transmitter and the relay node for the \emph{NFP} forwarding scheme $r_{N}$, we use the distribution of the distance to the nearest neighbour in Poisson networks given in \cite{distances}  with the condition of having at least one available node in the forward neighbourhood of the transmitter. Therefore, $f_{r_{N}}(r)$ can be obtained as follows:
\end{proof}
\end{lemma}

Finally, for the \emph{RFP} forwarding strategy, the transmitter-receiver distance is denoted by $r_{R}$ and its distribution can be expressed by the following lemma:
\begin{lemma}\label{rfp_dist}
The distance between the transmitter and a randomly selected relay inside the average transmission range of $\bar{R}_t$ can be expressed as follows:
\begin{equation}
f_{r_{R}} (r) = \frac{2 r}{\bar{R}_{t}^{2}}, \qquad \qquad \qquad \hfill{0 \leq r \leq \bar{R}_{t}.}
\end{equation}
\begin{proof}
See \textbf{Appendix~\ref{proof_rfp_dist}}.
\end{proof}
\end{lemma}
In the following section, we present the metric developed for performance analysis and provide the distributions of the forward packet progress.

%\begin{proof}
%For a uniformly random network, the cumulative distribution function (\emph{cdf}) is given by $F_{r_{R}}(r) = \mathbb{P}[r_{R} \leq r] =\pi r^{2}/ \pi R_{t}^{2}$. Differentiating the \emph{cdf} proves $f_{r_{R}}(r)$.
%\end{proof}

\section{Performance Characterization}\label{performance_analysis}
In this section, we study the performance of the depicted system model based on the three main metrics i.e. spatial frequency reuse, transmission quality and packet forward progress. The spatial frequency reuse is quantified by the intensity of simultaneously active transmitters i.e. the number of concurrent transmissions per unit area, which is given by $\Lambda(\rho_{th})$. The transmission quality is measured by the probability of transmission success $\mathcal{P}_{\chi}(\rho_{th})$. Finally, the data dissemination is governed by the forward packet progress at each hop.
The forward progress for different schemes $Z_{\chi},\chi \in \{M,N,R\}$ is a random variable and the \emph{pdf}s are given by the following lemma.
\begin{lemma} \label{fp_lemma}
The per-hop forward progress achieved by using the \emph{MFR} forwarding strategy in a fixed average transmission range $\bar{R}_t$ is distributed as follows:
\begin{align}
f_{{\rm Z}_{M}} ({\rm z}) &= \frac{2 \lambda \sqrt{\bar{R}_{t}^{2} - {\rm z}^{2}}}{1 - e^{- \lambda \pi \bar{R}_{t}^{2}/2}} e^{- \lambda \bar{R}_{t}^{2} \left( cos^{-1} (\frac{{\rm z}}{\bar{R}_{t}}) - (\frac{{\rm z}}{\bar{R}_{t}}) \sqrt{1 - ({\rm z}/\bar{R}_{t})^{2}} \right)}, \notag \\
& \qquad \qquad \qquad \qquad \qquad \qquad \qquad 0 \leq {\rm z} \leq \bar{R}_{t}.
\label{f_Z_M}
\end{align}
\normalsize
The per-hop forward progress achieved by using the \emph{NFP} forwarding strategy in a fixed average transmission range $\bar{R}_t$ is distributed as follows:
\begin{align}
f_{{\rm Z}_{N}} ({\rm z}) = \frac{\sqrt{2 \lambda} e^{- \lambda \pi {\rm z}^{2} /2} \rm{erf}\left(\sqrt{\frac{\pi \lambda}{2} (\bar{R}_{t}^{2} - {\rm z}^{2})} \right) }{1 - e^{- \lambda \pi \bar{R}_{t}^{2}/2}},
\hfill{0 \leq {\rm z} \leq \bar{R}_{t}}.
\label{f_Z_N}
\end{align}
\normalsize
The per-hop forward progress achieved by using the \emph{RFP} forwarding strategy in a fixed average transmission range $\bar{R}_t$ is distributed as follows:
\begin{align}
f_{{\rm Z}_{R}} ({\rm z}) = \frac{4 \sqrt{\bar{R}_{t}^{2} - {\rm z}^{2}}}{\pi \bar{R}_{t}^{2}}, \qquad \hfill{0 \leq {\rm z} \leq \bar{R}_{t}.}
\label{f_Z_R}
\end{align}
\begin{proof}
See \textbf{Appendix~\ref{proof_fp_lemma}}.
\end{proof}
\end{lemma}
\normalsize

%\begin{proof}
%Using \eqref{f_r_N} and the fact that the angle $\theta$ is independent and uniformly distributed between $-\pi/2$ and $\pi/2$, we can obtain $f_{r_{N},\theta}(r,\theta)$. Through transformation of variables from polar to Cartesian coordinates, we can find $f_{Z_{N},D_{N}}(z,d)$. Integrating over the range of $D_{N}$ proves the \emph{pdf} of $Z_{N}$.
%\end{proof}
%
%
%\begin{proof}
%Proof is similar to the proof of \eqref{f_Z_N} with $\chi = R$.
%\end{proof}

\begin{figure*}[t]
\centering
\subfloat[Probability of transmission success and spatial \newline reuse against carrier sensing threshold.]{\includegraphics[width = 2.4in]{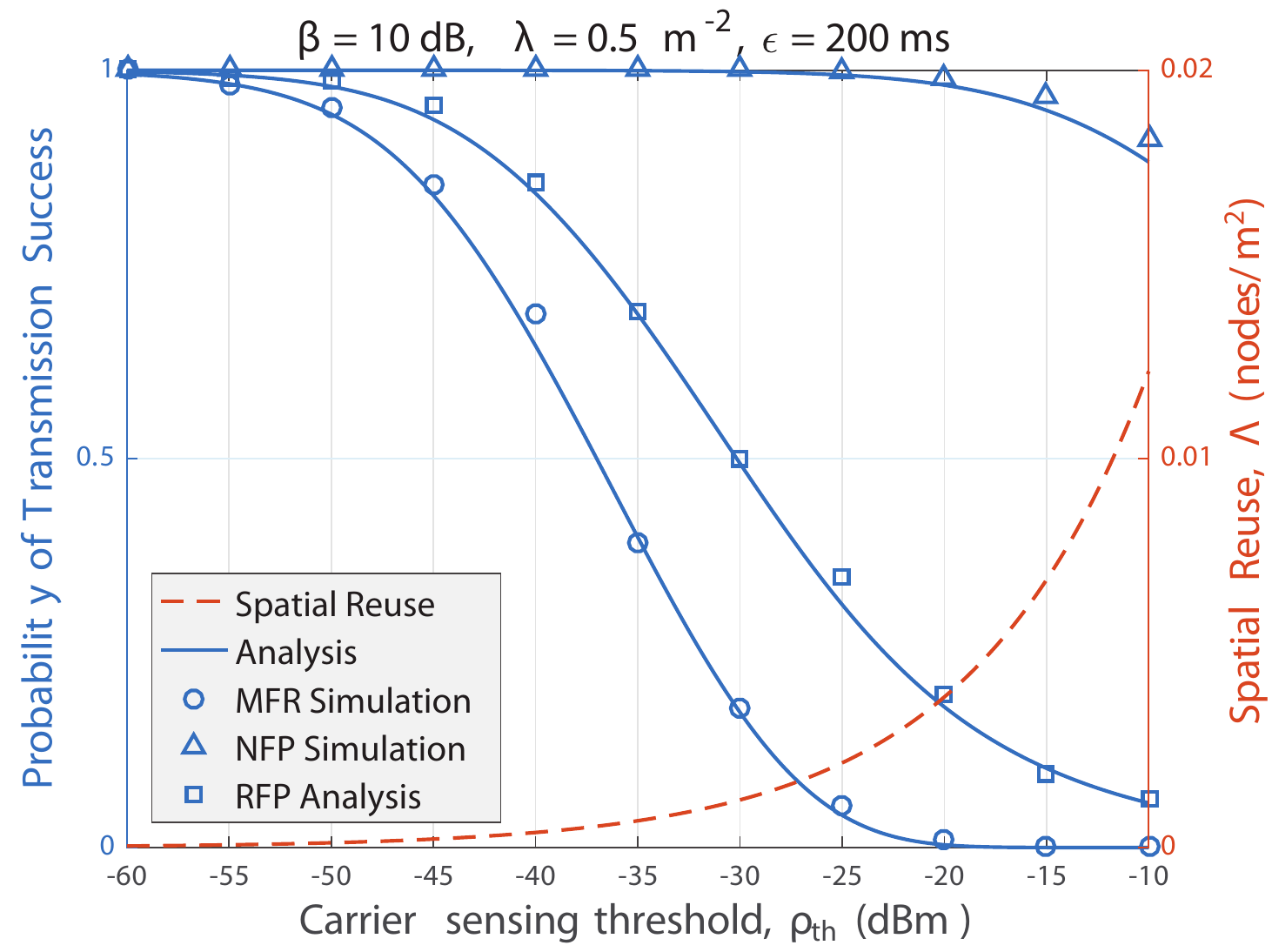} \label{P_s_fig}}
\subfloat[APP under low QoS requirements.]{\includegraphics[width = 2.4in]{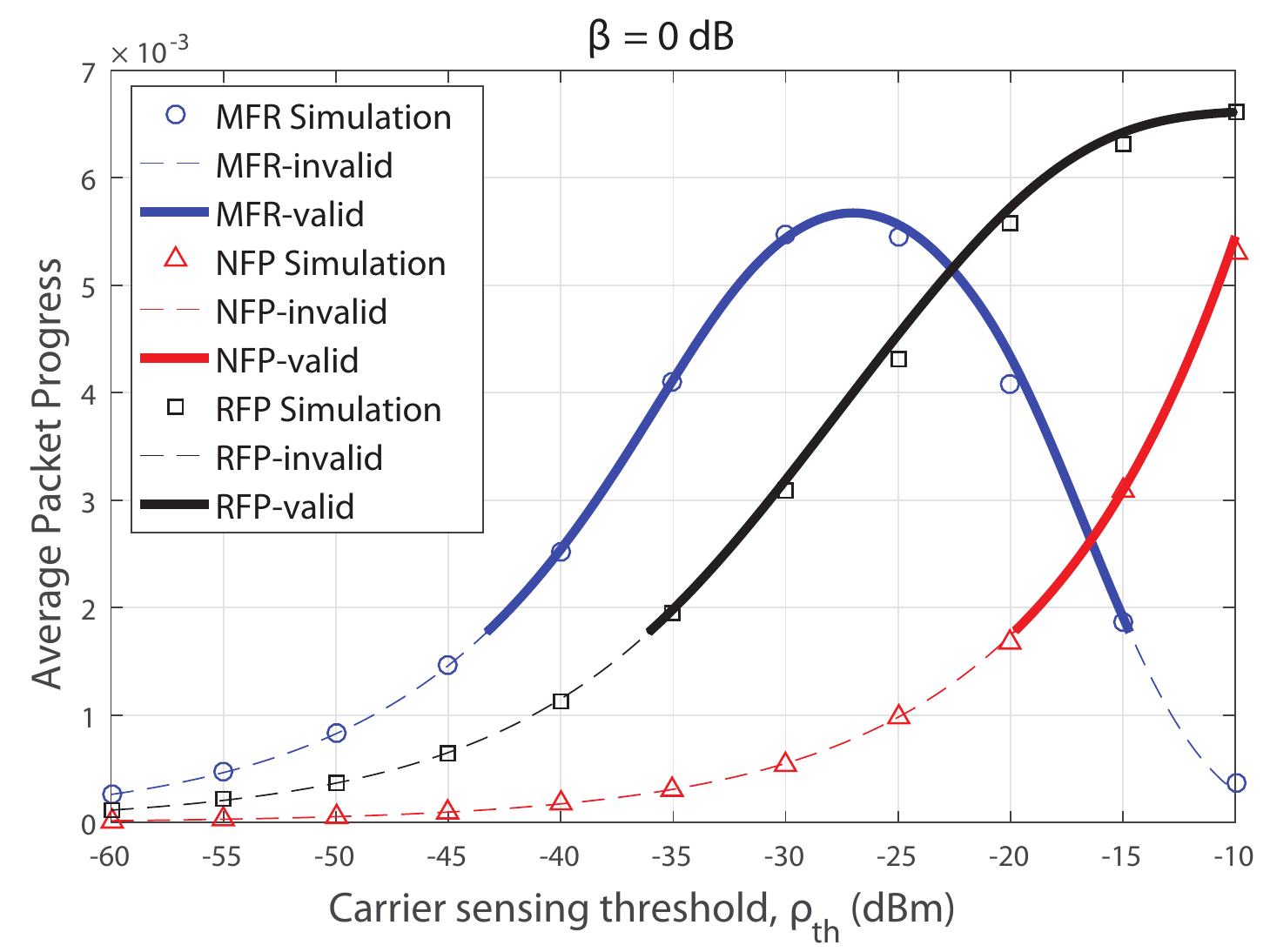} \label{small_delay}}
\subfloat[APP under high QoS requirements.]{\includegraphics[width = 2.4in]{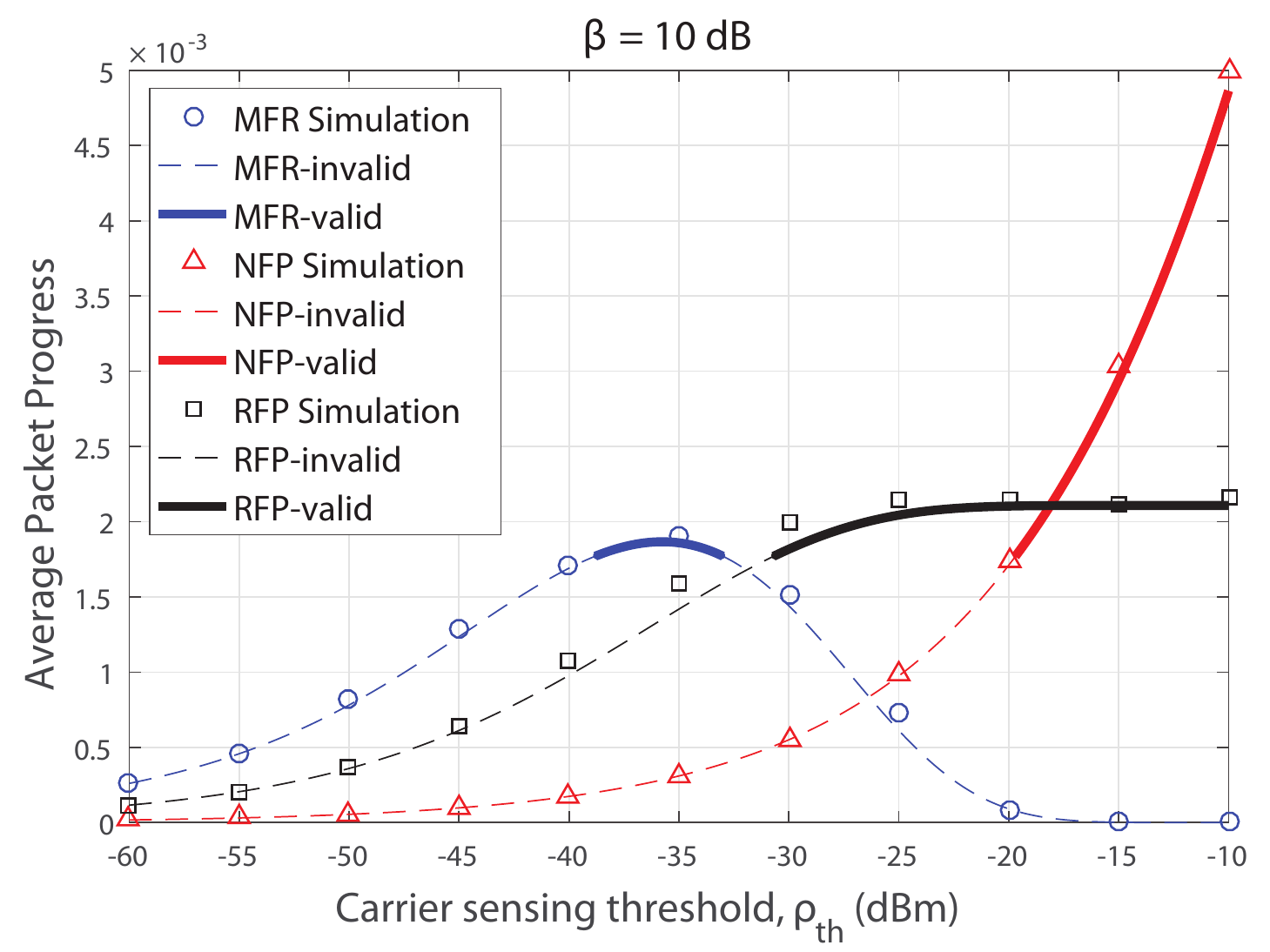} \label{big_delay}}
\caption{(a) The probability of transmission success is a decreasing function of the carrier sensing threshold as opposed to the spatial reuse, which is an increasing function. \emph{RFP} can achieve the highest APP under low transmission quality requirements while \emph{NFP} achieves the best APP performance under high transmission quality requirements with appropriate choice of $\rho_{th}.$}
\label{results}
\end{figure*}

The average per hop forward progress denoted by $\bar{Z}_{\chi} = \mathbb{E}[Z_{\chi}]$ for the three forwarding strategies can obtained using the \emph{pdf}s in \eqref{f_Z_M},\eqref{f_Z_N}, and \eqref{f_Z_R} for $\chi \in \{M,N,R\}$. Note that $\bar{Z}_{\chi}\sqrt{\lambda}$ can be conveniently used as a dimensionless measure of the forward progress and is therefore known as normalized average forward progress (NAFP) \cite{transmission_range_control}. In order to capture the effect of spatial frequency reuse, transmission quality and packet dissemination, we define a unified performance metric called the average packet progress (APP) similar to the one introduced in \cite{junaid_vehicular} as follows:
\begin{equation}
APP_{\chi}(\rho_{th}) = \mathcal{P}_{\chi}(\rho_{th}) \times \Lambda(\rho_{th}) \times \bar{{\rm Z}}_{\chi} \sqrt{\lambda},
\end{equation}
The physical meaning of APP is the average successful packet progress made per unit area in the network and the objective is to maximize the APP of the network. It is desirable as it maximizes each of the individual performance metrics. Note that for a fixed node density $\lambda$, the probability of transmission success depends on both $\chi$ and $\rho_{th}$, the density of simultaneous transmitters depends only on $\rho_{th}$ and the NAFP depends only on $\chi$. The challenge then is to jointly select the optimal values of $\chi$ and $\rho_{th}$ that achieve the best balance between the spatial reuse, transmission quality and the packet dissemination given specific throughput requirements. This problem can be formulated as:
\begin{equation}
\begin{matrix}
\displaystyle \max_{\chi,\rho_{th}} & APP_{\chi}(\rho_{th})\\
\textrm{s.t.} & \frac{\lambda}{\Lambda(\rho_{th})} \times \frac{1}{\mathcal{P}_{\chi}(\rho_{th})} \times \frac{ \delta}{ \bar{{\rm Z}}_{\chi}} \times \tau&\leq&\epsilon
\end{matrix}
\end{equation}

In this case, the throughput requirement comes from the maximum allowable message transfer delay that is determined by the mission critical nature of the application in the massive IoT network. The physical meaning of the constraint is that the total transmission delay from the source to the destination should not exceed a certain threshold $\epsilon$. The term $\frac{\lambda}{\Lambda(\rho_{th})}$ represents the average number of time slots required by a device to win access to the channel from its neighbors, $\frac{1}{P_{s}}$ signifies the average number of re-transmissions required to achieve successful per-hop packet transmission between the transmitter and receiver, $\frac{\delta}{\bar{Z}_{\chi}}$ represents the average number of hops that are required for the data to reach a destination device located at an arbitrary distance of $\delta$, and $\tau$ is the duration of each time slot. The goal is to find the best packet forwarding strategy and the optimal CST that maximizes the APP resulting in efficient data dissemination in mission critical massive IoT networks.

\vspace{-0.0in}
\section{Numerical Results}
\begin{figure}[]
  \centering
  \includegraphics[width=2.8in]{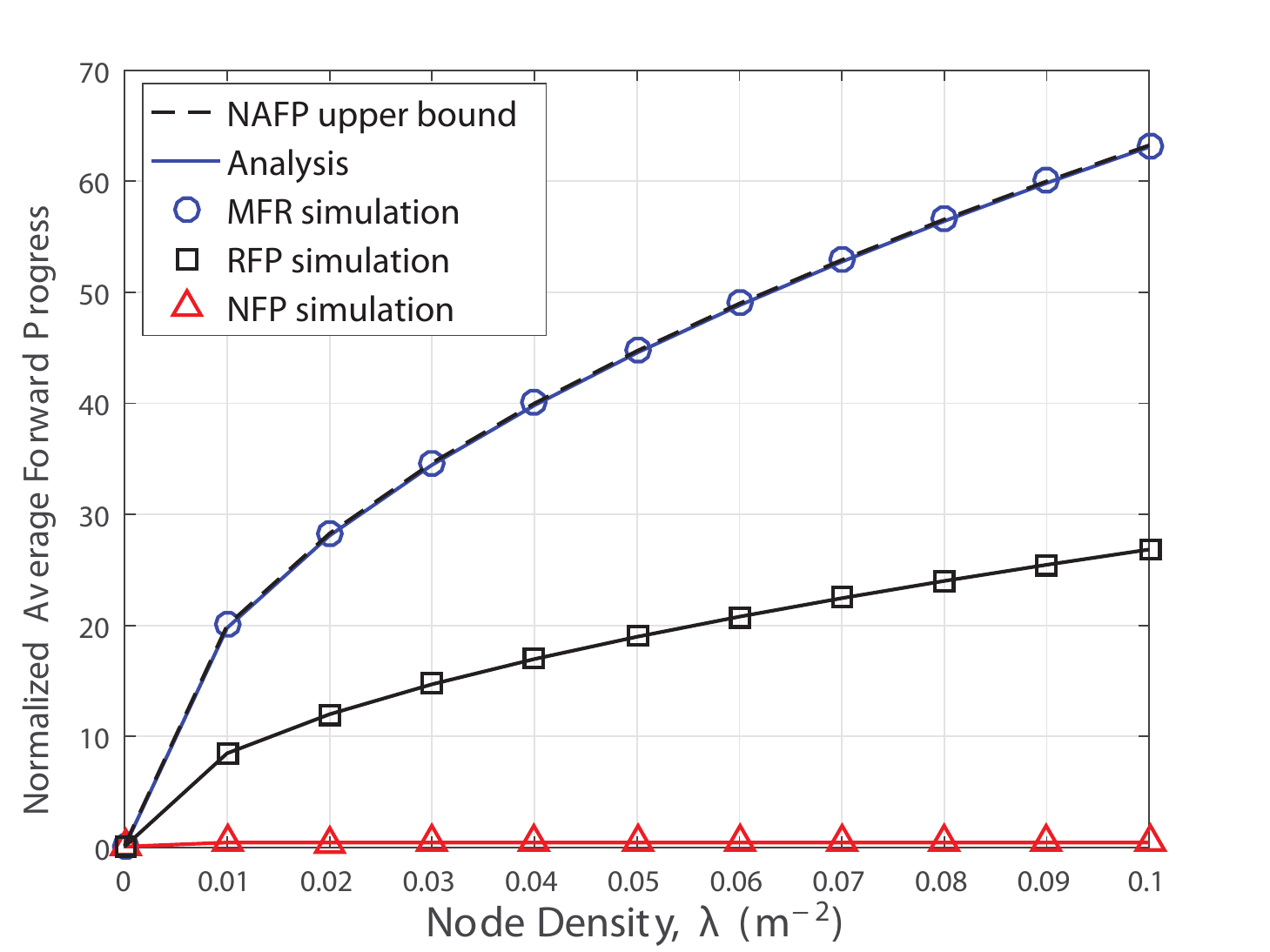}\\
  \caption{The \emph{MFR} scheme achieves the highest per hop NAFP followed by the \emph{RFP} and \emph{NFP} scheme for all node densities. The difference in NAFP between the \emph{NFP} and other schemes increases with the node density because of increasing number of available neighbours. The maximum possible NAFP is $R_{t}\sqrt{\lambda}$ and is therefore an upper bound.} \label{NAFP_fig}
\end{figure}

In this section, we will first use a numerical example to illustrate the different tradeoffs discussed in the paper. The choice of parameters is arbitrary and is made for illustrative purposes as follows: The transmit power $P = 0.1$ W, device density $\lambda = 0.5$ devices/${\rm m}^{2}$, transmission range $R_{t} = 10$ m (equivalently $\rho_{min} \approx -11$ dBm), path-loss exponent $\eta = 4$, SINR threshold $\beta = 10$ dB unless otherwise specified, distance to destination $\delta = 100$ m, maximum delay tolerance $\epsilon = 200$ ms and duration of time-slot $\tau = 10$ $\mu$s. Fig.~\ref{P_s_fig} shows the probability of transmission success $\mathcal{P}_{\chi}$ for the three packet forwarding schemes verified through Monte Carlo simulations. It is observed that increasing the CST reduces the probability of transmission success due to the reduced interference protection around the receiver and the increased number of concurrent transmissions. As expected, the \emph{NFP} scheme achieves the highest probability of success followed by \emph{RFP} and \emph{MFR} because of the increasing transmitter-receiver separation as shown in Fig~\ref{NAFP_fig}. Fig.~\ref{P_s_fig} also shows that the spatial reuse is an increasing function of the CST because increasing $\rho_{th}$ reduces the carrier sensing range and enables more nodes to qualify for transmission though the CSMA contention.

Fig.~\ref{small_delay} \& \ref{big_delay} show the performance of the network under the proposed APP metric for two different QoS requirements, i.e., $\beta = 0$ dB and $\beta = 10$ dB. We restrict the choice of CST to $-11$ dB so that the average sensing range remains greater than the average transmission range to guarantee that an interferer is not located arbitrarily close to the receiver for any of the forwarding strategies. Note that extremely high and extremely low CST results in high transmission delays. A low CST results in a high sensing range and therefore more contention, which leads to longer average time to sense an idle channel. On the other hand, a high CST results in a smaller sensing range and therefore lower interference protection, which results in more average number of retransmissions for successful message delivery. The region in which the average delay requirement is not met is represented by the dotted lines in Fig.~\ref{small_delay} \& \ref{big_delay} and is labeled as invalid. It can be observed that for low QoS requirements, the {\em RFP} strategy can achieve a higher APP performance, while for high QoS requirements, the NFP can lead to the highest APP with appropriate choice of CST.

In general, the choice of best packet forwarding strategy and CST depends on several factors such as the device density, QoS requirements, and the delay tolerance. However, the fact that the {\em RFP} achieves the highest APP performance indicates that there exists a more strategic forwarding strategy in the low QoS requirements case that can result in a further improvement in performance. However, if the QoS requirement is high, short hops via the  {\em NFP} strategy seems to be the best strategy.

%according to the proposed APP metric for the delay tolerance $\epsilon = 25$ ms and $\epsilon = 40$ ms. It is clear that with a relaxed delay constraint and an aggressive spectrum access, the \emph{NFP} forwarding scheme outperforms others to achieve the highest APP performance. However this is misleading because the unconstrained maximum leads to high transmission delays, and hence, lower throughput. Therefore, to conduct a fair comparison, the feasible region in which the constraint is satisfied is marked by bold lines. It can be observed that for smaller delay tolerance as in Fig.~\ref{small_delay}, the \emph{MFR} provides the best APP performance and the $\rho_{th}$ can be tuned to achieve the maximum APP. On the other hand, if the delay tolerance is relaxed as in Fig.~\ref{big_delay}, the \emph{NFP} is the best forwarding scheme with the appropriate choice of carrier sensing threshold.

\section{Conclusion}
This paper presents a framework for optimizing the mission critical data dissemination in massive IoT networks. The massive number of coexisting devices presents several performance challenges due to interference and packet delay over multi-hop transmissions. The performance tradeoffs are controlled by the carrier sensing threshold of the contention based spectrum access protocol as well as the packet forwarding strategy used at each hop. We develop an integrated framework that characterizes the performance of the system based on the spatial frequency reuse, probability of successful transmission, and the forward packet progress. A combined performance metric known as the average packet progress is developed that reflects the individual performance objectives. A delay constrained optimization problem is then presented which can be used to obtain the best balance between the performance objectives. The results show that the choice of forwarding scheme and the carrier sensing threshold, that optimize the network operation, depends on the delay tolerance and QoS requirements of the network.

%Future work can be directed towards investigating the performance improvements in massive IoT networks using other opportunistic variants of contention based spectrum access and/or more sophisticated packet forwarding strategies. Furthermore, a more generalized framework may be developed which takes into account the heterogeneity of IoT devices and their associated communication ranges.

\appendices

\section{Proof of Lemma~\ref{mfr_dist}} \label{proof_mfr_dist}
Using the approach in~\cite{transmission_range_control}, the relay that has the most forward progress has an empty segment farther than itself. The probability of such event can be used to obtain the joint distribution between $Z_{M}$ and $D_M$ as shown in Fig.~\ref{system_model}. By change of variables, i.e., $Z_{M} = r_M \text{cos} \theta$ and $D_{M} = r_M \text{sin} \theta$, the joint \emph{pdf} $f_{r_{M},\theta}(r, \theta)$ can be obtained. The marginal density for $r_{M}$ can then be obtained by integrating over $\theta$ from $-\frac{\pi}{2}$ to $\frac{\pi}{2}$ to obtain the result presented in Lemma~\ref{mfr_dist}. Note that the {\em pdf} has been conditioned on the event that there is at least one device in the forward neighbourhood of the transmitter.

\vspace{-0.1in}
\section{Proof of Lemma~\ref{rfp_dist}} \label{proof_rfp_dist}
The proof can be done using the cumulative distribution function (\emph{cdf}) approach. The \emph{cdf} of the random relay distance can be expressed as $F_{r_{R}}(r) = \mathbb{P}[r_{R} \leq r] =\pi r^{2}/ \pi \bar{R}_{t}^{2}$. Differentiating the \emph{cdf} with respect to $r$ gives $f_{r_{R}}(r)$ as shown in Lemma~\ref{rfp_dist}.

\vspace{-0.1in}
\section{Proof of Lemma~\ref{fp_lemma}} \label{proof_fp_lemma}
The {\em pdf} $f_{{\rm Z}_{M}}({\rm z})$ is obtained by integrating the joint density $f_{Z_M, D_M}(z,d)$, obtained in Appendix~\ref{proof_mfr_dist}, over the range of $D_M$. For deriving $f_{{\rm Z}_{N}}({\rm z})$, we can use \eqref{f_r_N} and the fact that the angle $\theta$ is independent and uniformly distributed between $-\pi/2$ and $\pi/2$ to obtain $f_{r_{N},\theta}(r,\theta)$. Through transformation of variables from polar to Cartesian coordinates, we can find $f_{{\rm Z}_{N},{\rm D}_{N}}(z,d)$. Integrating over the range of ${\rm D}_{N}$ proves the desired \emph{pdf} of $Z_{N}$. The {\em pdf} of $Z_R$, i.e., $f_{{\rm Z}_{R}}({\rm z})$ can be derived using a similar procedure as the {\em pdf} of $Z_N$.

\bibliographystyle{ieeetr}
\bibliography{references}

\end{document}